\documentclass[sigconf, nonacm]{acmart}

\newcommand\vldbyear{2026}
\newcommand\vldbworkshop{Tabular Data Analysis (TaDA)}
\newcommand\vldbauthors{\authors}
\newcommand\vldbtitle{\shorttitle}
\newcommand\vldbavailabilityurl{https://github.com/IBM/table-representation-evals}
\newcommand\vldbpagestyle{plain}

\newcommand\shorttitle{TEmBed-T}

\usepackage{multirow}
\usepackage{array}
\usepackage{makecell}
\usepackage{enumitem}
\usepackage{placeins}

\begin{document}

\newcommand{\todo}[1]{\textcolor{blue}{#1}}
\newcommand\benchmark{TEmBed}

\title[\benchmark-T: A Multi-Dimensional Benchmark for Table-Level Embeddings]{\benchmark-T: A Multi-Dimensional Benchmark\\for Table-Level Embeddings}

\author{Ayeen Poostforoushan}
\affiliation{%
  \institution{Independent Researcher}
}
\authornote{
Correspondence goes to \href{mailto:ayeen.pf@gmail.com}{ayeen.pf@gmail.com}}

\author{Liane Vogel}
\affiliation{%
  \institution{Technical University of Darmstadt}
}

\author{Carsten Binnig}
\affiliation{%
  \institution{Technical University of Darmstadt \& DFKI \& hessian.AI}
}

\begin{abstract}
Tabular data is the dominant structured-data modality, and learning table representations has become a core research direction. 
Table-level embeddings in particular underpin a wide range of applications, including table retrieval, data lake discovery, and table classification. 
Despite their importance, there is still limited understanding of how different embedding approaches behave across tasks, making systematic evaluation and analysis essential. 
In this work, we introduce a systematic evaluation of table-level embeddings that captures several complementary properties required for downstream effectiveness. 
We realize this evaluation by extending TEmBed, a recently proposed testbed for tabular embeddings, whose table-level coverage is currently limited to a single retrieval task.
An empirical study over the TEmBed model pool confirms that no single model excels across all tasks, demonstrating that table-level embedding quality cannot be reduced to retrieval alone.
\end{abstract}

\maketitle

\pagestyle{\vldbpagestyle}
\begingroup\small\noindent\raggedright\textbf{VLDB Workshop Reference Format:}\\
\vldbauthors. \vldbtitle. VLDB \vldbyear\ Workshop: \vldbworkshop.\\ 
\endgroup
\begingroup
\renewcommand\thefootnote{}\footnote{\noindent
This work is licensed under the Creative Commons BY-NC-ND 4.0 International License. Visit \url{https://creativecommons.org/licenses/by-nc-nd/4.0/} to view a copy of this license. For any use beyond those covered by this license, obtain permission by emailing \href{mailto:info@vldb.org}{info@vldb.org}. Copyright is held by the owner/author(s). Publication rights licensed to the VLDB Endowment. \\
\raggedright Proceedings of the VLDB Endowment. 
ISSN 2150-8097. \\
}\addtocounter{footnote}{-1}\endgroup

\ifdefempty{\vldbavailabilityurl}{}{
\vspace{.3cm}
\begingroup\small\noindent\raggedright\textbf{VLDB Workshop Artifact Availability:}\\
The source code, data, and other artifacts have been made available at \url{\vldbavailabilityurl}.
\endgroup
}

\section{Introduction}
\label{sec:introduction}

\noindent\textbf{The Importance of Table-Level Embeddings.}
Tabular data is the dominant modality in databases, enterprise systems, and the open web. Learning embeddings---that is, vector representations---of tables has therefore become a central challenge at the intersection of databases and machine learning. Although a growing number of table encoders have been proposed, systematic methods for evaluating and comparing the representations they produce remain underdeveloped \cite{badaro2023trl_survey}. Table-level embeddings are particularly important because they support retrieval, data discovery, schema understanding, and downstream prediction at the scale of entire tables. Yet, despite their relevance to these applications, relatively few models are explicitly designed to produce general-purpose table-level representations. This gap makes it difficult to determine which models are best suited to different table-centric workloads.

\noindent\textbf{Limitations of Existing Benchmarks.}
Existing benchmarks either measure end-to-end task performance, such as retrieval \cite{ji2025target_benchmark}, question answering \cite{ji2025target_benchmark}, and joinability detection \cite{srinivas2023lakebench}, or probe specific invariances of tabular structures \cite{cong2023observatory}. Consequently, they often conflate representation quality with downstream task success and provide limited insight into the capabilities that drive performance.
We therefore argue that meaningful evaluation of table-level embeddings requires disentangling these capabilities and assessing them independently. Different downstream applications rely on different embedding properties, and models that perform well along one dimension may perform poorly along another. Property-level evaluation thus enables more meaningful comparisons and supports more informed model selection.
\benchmark{} \cite{tembed_benchmark} takes an important step toward unified evaluation  \cite{muennighoff2023mteb} across four embedding granularities: cell, row, column, and table. At the table level, however, its evaluation remains limited and does not isolate the properties that embeddings actually capture. 

\noindent\textbf{Our Contribution: \benchmark{}-T.}
To address these limitations, we extend \benchmark{}'s table-level evaluation with \benchmark{}-T. We introduce three additional tasks, each designed to assess a distinct and practically relevant capability. First, we extend retrieval to the query-to-table setting and evaluate it across seven heterogeneous corpora \cite{ji2025target_benchmark}. This task measures cross-domain semantic alignment and complements \benchmark{}'s existing table-to-table setup, thereby covering both query-to-table and table-to-table retrieval scenarios encountered in practice. Second, we introduce table shuffling, a controlled setting that disrupts relational structure while preserving surface content. This task assesses whether embeddings capture structural relationships rather than relying primarily on semantic or lexical cues. Third, we add table type detection \cite{chen2023hytrel} on header-stripped tables, isolating the ability to infer table semantics from cell values alone while removing reliance on schema information.

\noindent\textbf{Key Findings.}
Across five representative approaches, we find that model rankings diverge substantially across these tasks. Importantly, as we show in our initial evaluation, no single embedding approach performs best across all evaluated properties, and strong performance in one setting does not reliably transfer to others. These findings underscore the need for property-driven evaluation and position \benchmark{}-T as a diagnostic complement to \benchmark{}'s existing benchmark.

\section{\benchmark{}-T: Benchmark Extensions}
\label{sec:tasks}

We identify three relevant properties based on which we select the tasks to extend \benchmark{}. We first discuss these properties, before we present our extensions in Section~\ref{sec:table-retrieval} to \ref{sec:table-type-detection}:

\noindent\textbf{Cross-domain Robustness.} is important as many applications like NL2SQL or table QA need to be able to work with data from various domains.
An embedder that silos search to one domain is not usable at scale. We probe this property with table retrieval across seven corpora spanning diverse origins, schemas, and sizes (Section~\ref{sec:table-retrieval}).

\noindent\textbf{Structural Fidelity.} defines whether the encoder preserves the structure of a table. 
Consider a table of CEOs and companies:
\begin{center}
\begin{minipage}[t]{0.46\columnwidth}
\centering
\footnotesize
\begin{tabular}{ll}
\hline
CEO & Company \\
\hline
Elon Musk & Tesla \\
Andy Jassy & Amazon \\
\hline
\end{tabular}\\[2pt]
{\footnotesize Anchor}
\end{minipage}
\hfill
\begin{minipage}[t]{0.46\columnwidth}
\centering
\footnotesize
\begin{tabular}{ll}
\hline
CEO & Company \\
\hline
Elon Musk & Amazon \\
Andy Jassy & Tesla \\
\hline
\end{tabular}\\[2pt]
{\footnotesize Value-shuffled (negative)}
\end{minipage}
\end{center}
Permuting rows and columns changes only surface order while preserving row-wise integrity; a structure-aware embedder should produce a near-identical vector.\footnote{The permutation-based positive assumes that row and column order carry no intrinsic semantics. For tables where ordering is meaningful (e.g., temporally sorted tables), this assumption may not hold.} Shuffling values independently within columns, however, scrambles those associations: the value-shuffled table destroys row integrity while preserving the exact multiset of tokens. A bag-of-words model or an encoder whose attention ignores table structure cannot distinguish these two transformations.
We operationalize this property with the table shuffling triplet protocol (Section~\ref{sec:table-shuffling}).

\noindent\textbf{Header-independent Semantic Preservation.} captures whether the embedding retains the table's high-level identity from cell content alone. An embedder that lacks this property can be misled by a surface-level header: a table whose header reads \texttt{Event} but whose cells describe a restaurant may be embedded closer to event tables simply because the header token dominates the representation. Stripping headers forces the embedding to recover type from cell values, isolating whether the model abstracts over data rather than memorizing header vocabulary. We evaluate this property with the task of table type detection (Section~\ref{sec:table-type-detection}).

\subsection{Table Retrieval}
\label{sec:table-retrieval}

Retrieving relevant tables e.g. out of a data lake given a query is usually performed by embedding each table and ranking the candidate tables in the corpus by relevance to the query embedding.
The task is a reasonable choice to probe cross-domain robustness of table embeddings, as a robust embedding model must hold its ranking quality across heterogeneous corpora out of multiple domains. 

\noindent\textbf{Design Dimensions.}
Our framework lets users inspect embeddings by changing different dimensions of the experiments:

\noindent\emph{Row count.} 
For fine-grained analysis, the row count included in table embeddings can be set from zero (table headers only), to using full table sizes. For this paper we report results with headers only (zero rows) and with 100 rows per table.

\noindent\emph{Serialization.} 
We test markdown and CSV serialization to isolate sensitivity to surface formatting independently of content. At headers-only the two formats collapse to near-identical strings, so this axis matters only with rows present. Serialization applies only to text-based encoders.

\noindent\textbf{Dataset.} We evaluate over the five corpora packaged by TARGET~\cite{ji2025target_benchmark}.
FeTaQA \cite{fetaqa}, TabFact \cite{tabfact}, and OTT-QA\cite{ottqa} are Wikipedia-derived corpora that pair tables with free-form QA, fact verification, and multi-hop QA queries respectively, covering general open-domain table content. Spider\footnote{Spider comes separated into train, validation and test splits, we use all three of them for evaluation}\cite{spider}, and BIRD \cite{bird} are NL2SQL corpora with relational databases covering various domains.
Together the corpora span individual tables and relational
schemas, and varying corpus sizes, exercising heterogeneity that a single-corpus benchmark cannot.

\vspace{-0.5em}
\subsection{Table Shuffling}
\label{sec:table-shuffling}

We test structural fidelity with a triplet protocol. For an anchor table $T$, we construct a structurally faithful positive $T^{+}$ and a value-shuffled negative $T^{-}$, and evaluate their embedding distances to the anchor:
\qquad $d_{\mathrm{pos}} = d_{\mathrm{cos}}(f(T), f(T^{+})), \quad d_{\mathrm{neg}} = d_{\mathrm{cos}}(f(T), f(T^{-})).$

$T^{+}$ is obtained by permuting rows and columns, which preserves row-wise associations and thus relational structure.
$T^{-}$ shuffles values independently within each column, breaking row-wise associations while preserving the exact token multiset. 
Since both variants share identical token distributions, any separation must arise from structural encoding rather than lexical cues.

\noindent\textbf{Design Dimensions.} The triplet generation is controlled by four parameters that together determine discrimination difficulty:

\noindent\emph{Permutation type.} Reordering rows and/or columns determines which structural axis is disrupted. Comparing row-against column-only accuracy reveals encoder-specific directional sensitivity.

\noindent\emph{Positive perturbation magnitude.} $d_{+} \in [0,1]$ is the fraction of rows/ columns randomly pairwise-swapped to generate $T^{+}$ while preserving row-wise associations. 

\noindent\emph{Negative perturbation magnitude.} $f_{-} \in [0,1]$ is the fraction of columns selected for intra-column shuffling, and $d_{-} \in [0,1]$ is the fraction of rows swapped within each selected column. 

\noindent\emph{Table size ablation.} Tables are filtered by row and column bounds before triplet generation. Three windows define the size axis: default ($\leq 100 \times 30$), BIG ($\leq 500 \times 80$), and SMALL ($\leq 20 \times 8$), all at fixed $d_{+} = 0.75,\ f_{-} = 0.7,\ d_{-} = 0.8$. Performance differences isolate whether structural perception depends on table dimensions.

\noindent\textbf{Datasets.} We generate triplets from four TARGET datasets~\cite{ji2025target_benchmark} containing semantically understandable cell values and from CKAN and ECB~\cite{srinivas2023lakebench} with mostly statistical and numerical data.

\subsection{Table Type Detection}
\label{sec:table-type-detection}

Table type detection (TTD) evaluates header-independent semantic preservation \cite{chen2023hytrel}.
Given a header-stripped table, we train a probe classifier on frozen embeddings to predict the table's Schema.org type from cell content alone.
Unlike HyTrel \cite{chen2023hytrel}, which fine-tunes the encoder end-to-end, we fix the encoder and treat classification performance as a measure of retained task-information information, analogous to row-level probing in \benchmark{} \cite{tembed_benchmark}.
Removing headers prevents trivial reliance on header tokens, forcing the model to capture schema-level semantics from cell values.

\noindent\textbf{Design Dimensions.} Two parameters control the protocol:

\noindent\emph{Classifier.} Three probe classifiers (XGBoost\cite{chen2016xgboost}, MLP, and KNN) are trained independently on the same frozen embeddings. Consistent rankings across classifiers indicate that recoverability is driven by the embedding rather than the classifier.  

\noindent\emph{Serialization.} We evaluate both markdown and CSV serializations (Section \ref{sec:table-retrieval}).

\noindent\textbf{Dataset.} We use the TTD dataset from HyTrel \cite{chen2023hytrel}, which is based on the WDC Schema.org corpus \cite{wdc-schemaorg}.
For efficiency and balanced macro-F1 evaluation, we subsample 800 training and 100 test tables for each of the 10 classes (8,000/1,000 total).

\newpage
\section{Initial Evaluation}
\label{sec:experiments}

\subsection{Setup}

\noindent\textbf{Models.}
\label{sec:approach-pool}
\noindent
We evaluate on the four approaches included in \benchmark~ that produce table level embeddings. In addition, we add a term-frequency baseline as the extreme case of structural blindness in the table shuffling task and as a naive baseline across all tasks.
The model pool spans three families: text-serialization transformers (MiniLM \cite{wang2020minilm}, Granite-R2 \cite{awasthy2025granite-r2}, GritLM \cite{muennighoff2024gritlm}), a structure-aware table encoder (HyTrel \cite{chen2023hytrel}), and the term-frequency baseline (Hashing).
For the Hashing approach, we use scikit-learn's \cite{scikit-learn} \texttt{HashingVectorizer}, which maps tokens to a fixed-size sparse hashed term-frequency vector. Token counts are accumulated in hash buckets and L2-normalized, producing a lexical-only embedding. We use embedding dimension of $32,768$ in this benchmark.

\noindent
\noindent\textbf{Metrics.}
\label{sec:metrics}
\noindent\textit{Table Retrieval.} We report MRR and Recall on each corpus.

\noindent\textit{Table Shuffling.} We quantify the triplet condition $d_{\mathrm{pos}} < d_{\mathrm{neg}}$ (Section~\ref{sec:table-shuffling}) with two complementary metrics.
 Triplet accuracy is the probability that the positive is closer to the anchor than the negative; scores below $0.50$ indicate systematic inversion. The silhouette score measures the strength of that classification margin.
\begin{equation}
\mathrm{Silhouette} = \mathbb{E}\!\left[\frac{d_{\mathrm{neg}} - d_{\mathrm{pos}}}{\max(d_{\mathrm{pos}},\; d_{\mathrm{neg}})}\right] \in [-1, 1]
\end{equation}

\noindent\textit{Table Type Detection.} We report F1-macro 
across all three classifiers.

\noindent
\noindent\textbf{Statistical Significance.}
\label{sec:bootstrap-ci}
\noindent
For Table Retrieval and Table Shuffling, we report bootstrapped 95\% confidence intervals obtained by resampling per-instance scores with replacement (per-query for retrieval, per-triplet for shuffling; 10,000 resamples).

\subsection{Table Retrieval}
\label{sec:retrieval-results}

\begin{figure}
\centering
\includegraphics[width=\columnwidth]{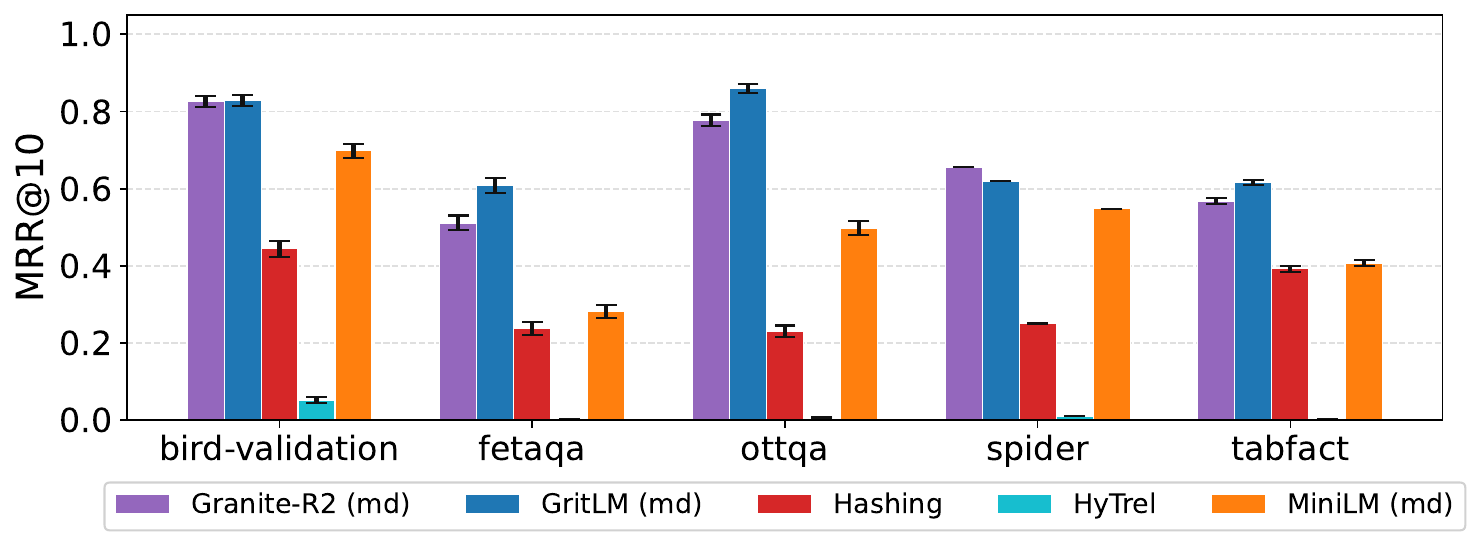}
\caption{Table Retrieval: MRR@10 per dataset at $\texttt{row\_limit}=100$. Spider: aggregated over its three splits. Dataset heterogeneity drives domain-specific ranking reversals between leading approaches. Error bars show bootstrapped 95\% CIs.}
\label{fig:per-dataset-bars}
\end{figure}

We evaluate table retrieval with headers-only and the first 100 rows, as well as markdown and CSV serialization, as described in Section~\ref{sec:table-retrieval}.
Figure~\ref{fig:per-dataset-bars} shows per-dataset MRR@10 at $\texttt{row\_limit}=100$.
Dense transformer encoders lead retrieval across most corpora, reflecting pretraining on natural-language text that transfers to table cell and header vocabulary.
Domain-specific reversals emerge within this group: GritLM performs best on Wikipedia-derived corpora (FeTaQA, TabFact, OTT-QA), while Granite-R2 surpasses the other transformers on Spider.
This reversal reflects Granite-R2's training emphasis on relational table data \cite{awasthy2025granite-r2}.
The cross-domain divergence confirms that no single encoder generalizes uniformly, and that evaluating over heterogeneous corpora is necessary to expose training-distribution bias.

\begin{figure}
\centering
\includegraphics[width=\columnwidth]{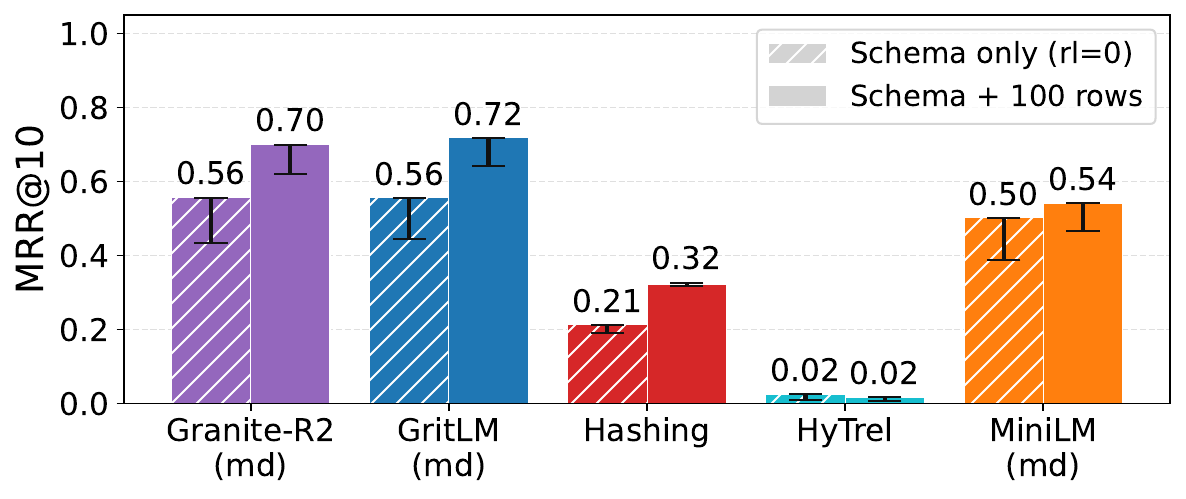}
\caption{Table Retrieval: MRR@10 at $\texttt{row\_limit}=0$ (headers only) vs.\ $\texttt{row\_limit}=100$, averaged over datasets. The gap between bars reveals cell content leverage per approach. Error bars show bootstrapped 95\% CIs.}
\label{fig:row-limit-bars}
\vspace{-1em}
\end{figure}

Figure~\ref{fig:row-limit-bars} compares retrieval performance under schema-only and first-100-rows configurations, averaged over datasets.
All approaches improve when row content is provided, confirming that cell values carry signal beyond headers.
All transformer encoders gain substantially from row content, while MiniLM shows a markedly smaller gain. This asymmetry indicates that MiniLM's lower-capacity representation already saturates on schema-level signal, leaving little room to leverage additional cell content.
The comparison of serialization formats is reported in Table~\ref{tab:retrieval_md_csv} in the Appendix.

\subsection{Table Shuffling}
\label{sec:shuffling-results}

We evaluate across 15 variations crossing permutation type, magnitude regime, and table window size; the full grid and ablations are detailed in Appendix~\ref{sec:shuffling-ablations}. Figure \ref{fig:headline-bars} (middle) shows that all models except Hytrel perform poorly on the task.
The full per-dataset accuracy breakdown at the canonical v0 variation (Table~\ref{tab:shuffling_accuracy}, Appendix) confirms that the structure-aware encoder consistently distinguishes structurally faithful from value-shuffled tables across all datasets. Transformer-based approaches and the bag-of-words baseline both score at or below random chance on lexically rich datasets, but the silhouette score (Figure~\ref{fig:silhouette-bars}, Appendix~\ref{sec:shuffling-ablations}) exposes two distinct failure modes underneath: Hashing ties exactly at zero, since its permutation-invariant nature cannot represent the anchor, positive, and negative as different vectors in the first place; transformers instead score negative since their attention tracking surface token order over tabular structure, so the whole-row/column permutation behind the positive disturbs that order more than the localized value-swaps behind the negative.

The ECB dataset provides the clearest diagnostic case. Its statistical content carries minimal lexical discriminability, so any positive-versus-negative separation must arise from structural encoding rather than lexical cues. Transformer-based approaches and the bag-of-words baseline collapse entirely on ECB with 0\% accuracy, consistent with the distinct failure modes above. The structure-aware encoder maintains near-perfect accuracy, directly isolating the structural contribution by eliminating the lexical shortcut.

Figure~\ref{fig:row-col-scatter} (Appendix~\ref{sec:shuffling-ablations}) reveals encoder-specific directional asymmetries: row reordering and column reordering yield distinct accuracy profiles per approach, indicating that different encoders represent the two structural axes differently.

\begin{figure*}[t]
\centering
\includegraphics[width=0.95\textwidth]{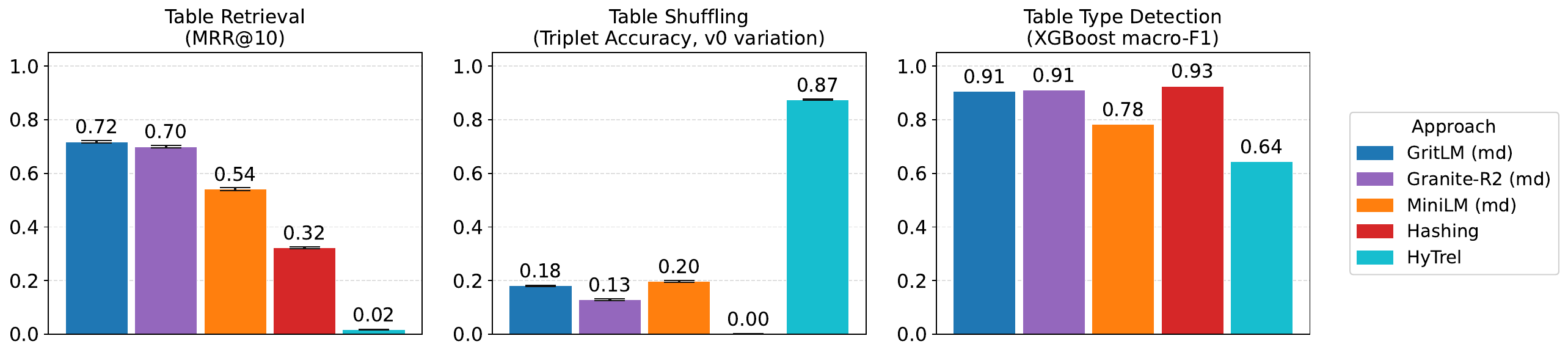}
\vspace{-1.5em}
\caption{Per-task scores for all five approaches, averaged over each task's datasets. Markdown serialization was used for text-serialization transformers. For each task a different model works best: GritLM for Retrieval, HyTrel for Shuffling, Hashing for TTD. Error bars on the Retrieval and Shuffling panels show bootstrapped 95\% CIs.
}
\label{fig:headline-bars}
\end{figure*}

\subsection{Table Type Detection}
\label{sec:ttd-results}

\begin{figure}
\centering
\includegraphics[width=\columnwidth]{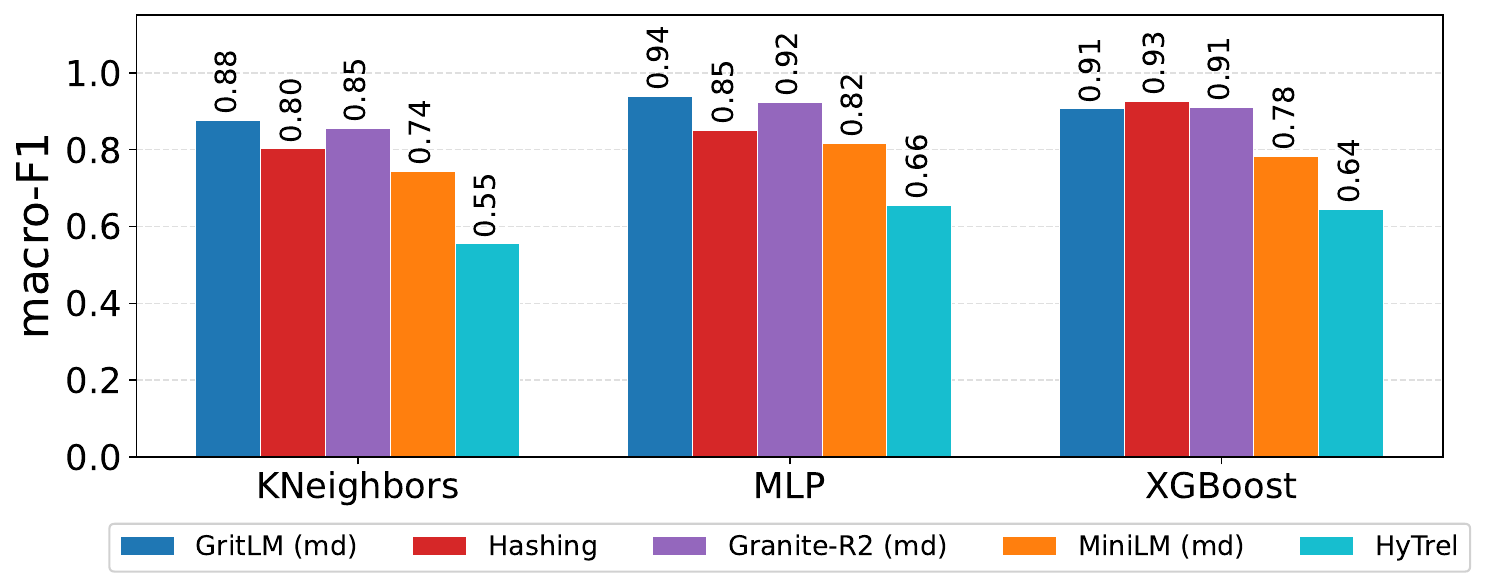}
\vspace{-2em}
\caption{Table Type Detection: Macro-F1 per approach and classifier on header-stripped WDC Schema.org tables. Markdown-serialization was used for the text-serialization transformers.}
\label{fig:ttd-classifier-bars}
\vspace{-1.5em}
\end{figure}

We train three probe classifiers (XGBoost \cite{chen2016xgboost}, MLP, KNN) on frozen embeddings and evaluate at markdown and CSV serialization, as described in Section~\ref{sec:table-type-detection}.
Figure~\ref{fig:ttd-classifier-bars} reports macro-F1 across classifiers. Classifier choice affects absolute scores but not approach rankings, which are consistent across probe types. The term-frequency baseline (Hashing) leads the tree-based classifier, confirming that table type is recoverable from surface token patterns alone. Dense transformer approaches achieve stronger performance under neural probes, indicating their representations compress type-relevant semantic structure beyond what token frequencies encode. The structure-aware encoder underperforms across all classifiers, consistent with pretraining objectives that optimize structural relationships rather than schema-type signal. Complete per-classifier results are in Table~\ref{tab:ttd_classifier} (Appendix).


\subsection{Cross-Task Overview}
\label{sec:cross-task-overview}

Table~\ref{tab:overall_ranking} (Appendix~\ref{sec:cross-task-results}) reports per-task ranks and an average rank across the three tasks. Rankings diverge sharply across tasks, with no single encoder leading on all three axes. This directly motivates the need for models with broad, multi-dimensional table-level capabilities that do not yet exist.
The comparison of CSV and markdown serialization is shown in Figure~\ref{fig:serialization-deltas} (Appendix~\ref{sec:cross-task-results}).
The results vary for each approach and task combination, confirming that there is no universally superior format between CSV and markdown.
In addition to measuring performance, we also measured execution time, as visualized in Figure~\ref{fig:cost-quality} (Appendix~\ref{sec:cross-task-results}).
A general trend holds where higher processing cost yields better pooled quality, with Hashing as the exception since its overall quality is poor despite high embedding cost, consistent with its nature as a classical lexical-only baseline. Granite-R2 achieves strong pooled quality despite being substantially smaller than GritLM, owing to its training on a large corpus of relational table data~\cite{awasthy2025granite-r2}.

\section{Related Work}
\label{sec:related-work}

TARGET~\cite{ji2025target_benchmark} benchmarks retrieval for generative tasks such as question answering, fact verification, and text-to-SQL. LakeBench~\cite{srinivas2023lakebench} evaluates data lake discovery across multiple tasks.
\noindent Observatory~\cite{cong2023observatory} takes a diagnostic approach, defining eight primitive properties grounded in relational invariants and data distribution considerations. It measures how embeddings respond to controlled perturbations at the row and column level, quantifying embedding drift across the shuffles. \noindent \citet{younes2026unified} similarly focus on robustness to benign perturbations, applying row and column permutations and measuring the similarity between the original and perturbed embeddings.
Both frameworks confirm that encoders vary in their sensitivity to reordering. In a permutation probe, however, the anchor and the permuted table share the exact same multiset of tokens, so an encoder operating as a bag-of-words model passes by matching token overlap without encoding structural information.
Our value-shuffled negative fills this gap by preserving each column's token multiset while scrambling cross-column associations, so any separation must arise from structural encoding alone.

\section{Conclusion}
\label{sec:conclusion}
\benchmark{}'s \cite{tembed_benchmark} table-level evaluation axis is limited to a single retrieval task, leaving the intrinsic properties a table encoder must possess entirely untested. We extend it into a three-task diagnostic suite grounded in both the fundamental characteristics a table embedder needs to have and the application-level properties it is expected to deliver in practice. A model that performs well across all three axes can be considered a genuinely capable table-level encoder.
We evaluated the suite through a range of experiments and ablation sweeps that demonstrate the benchmark's ability to probe encoders along independent diagnostic axes. Per-task rankings diverge substantially across all approaches, and no current model performs well on all three axes simultaneously. This confirms that table-level embedding quality is multi-dimensional, and building an encoder that covers the full capability profile remains an open problem.

\begin{acks}
This work was funded by the BMBF and the state of Hesse as part of the NHR program, by the LOEWE Spitzenprofessur (III 5-519/05.00.003-(0005)), by the Deutsche Forschungsgemeinschaft (DFG, German Research Foundation), and under Germany’s Excellence Strategy (EXC-3057/1 “Reasonable Artificial Intelligence”, Project No. 533677015). We also thank DFKI and hessian.AI.
\end{acks}

\bibliographystyle{ACM-Reference-Format}
\bibliography{references}

\FloatBarrier
\appendix
\section{Appendix}
\label{sec:appendix}

This appendix provides full experimental detail supporting the results summarized in Section \ref{sec:experiments}: per-dataset and per-serialization breakdowns for Table Retrieval (\ref{sec:appendix-retrieval}), the complete variation grid and ablations for Table Shuffling (\ref{sec:shuffling-ablations}), per-classifier results for Table Type Detection (\ref{sec:appendix-ttd}), and cross-task comparisons of serialization, ranking, and cost-quality trade-offs (\ref{sec:cross-task-results}).
\subsection{Table Retrieval}
\label{sec:appendix-retrieval}

While Figure~\ref{fig:per-dataset-bars} reports the results using markdown serialization for Granite-R2~\cite{awasthy2025granite-r2}, GritLM~\cite{muennighoff2024gritlm}, and MiniLM~\cite{wang2020minilm}, we report the full results of the format ablations per dataset in Table~\ref{tab:retrieval_md_csv}.
Results vary by dataset, but markdown tends to achieve higher results overall, especially for GritLM, whereas MiniLM scores slightly higher on average with CSV serialization.

\begin{table*}[t]
\centering
\setlength{\tabcolsep}{8.6pt}
\begin{tabular*}{\textwidth}{lccccc|c}
\hline
Approach & bird-validation & fetaqa & ottqa & spider-train & tabfact & Mean \\
\hline
Granite-R2 (csv) & 0.8231 {\scriptsize [0.809, 0.837]} & 0.5015 {\scriptsize [0.482, 0.521]} & 0.7675 {\scriptsize [0.753, 0.782]} & \textbf{0.6090} {\scriptsize [0.599, 0.619]} & 0.5500 {\scriptsize [0.542, 0.558]} & 0.6502 \\
Granite-R2 (md) & \underline{0.8257} {\scriptsize [0.811, 0.840]} & 0.5112 {\scriptsize [0.492, 0.531]} & 0.7777 {\scriptsize [0.763, 0.792]} & \underline{0.5932} {\scriptsize [0.583, 0.603]} & 0.5673 {\scriptsize [0.560, 0.575]} & 0.6550 \\
GritLM (csv) & 0.7810 {\scriptsize [0.765, 0.797]} & \textbf{0.6196} {\scriptsize [0.601, 0.638]} & \underline{0.8569} {\scriptsize [0.845, 0.868]} & 0.5219 {\scriptsize [0.512, 0.532]} & \textbf{0.6183} {\scriptsize [0.611, 0.626]} & \underline{0.6795} \\
GritLM (md) & \textbf{0.8290} {\scriptsize [0.814, 0.843]} & \underline{0.6086} {\scriptsize [0.590, 0.627]} & \textbf{0.8605} {\scriptsize [0.849, 0.872]} & 0.5527 {\scriptsize [0.543, 0.562]} & \underline{0.6161} {\scriptsize [0.609, 0.624]} & \textbf{0.6934} \\
MiniLM (csv) & 0.6913 {\scriptsize [0.673, 0.709]} & 0.2980 {\scriptsize [0.280, 0.316]} & 0.5086 {\scriptsize [0.491, 0.526]} & 0.4942 {\scriptsize [0.484, 0.504]} & 0.4291 {\scriptsize [0.422, 0.437]} & 0.4842 \\
MiniLM (md) & 0.6986 {\scriptsize [0.680, 0.716]} & 0.2819 {\scriptsize [0.264, 0.299]} & 0.4983 {\scriptsize [0.481, 0.516]} & 0.4825 {\scriptsize [0.473, 0.492]} & 0.4072 {\scriptsize [0.400, 0.415]} & 0.4737 \\
\hline
\end{tabular*}
\caption{Table Retrieval MRR@10: markdown vs CSV serialization (rl=100, transformers only; spider validation and test splits showed near-identical trends and are omitted for space). Bracketed values show bootstrapped 95\% CIs.}
\label{tab:retrieval_md_csv}
\end{table*}

\subsection{Table Shuffling: Variation Grid \& Ablations}
\label{sec:shuffling-ablations}

The Table Shuffling evaluation is structured as a grid of 15 variations across three axes: permutation type (both, row reorder, column reorder), perturbation magnitude, and table window size, as described in Section~\ref{sec:table-shuffling}.

\noindent\textbf{Variation grid.}
Table~\ref{tab:shuffling-grid} lays out the main magnitude grid (v0--v8).
For high positive perturbation we use $d_{+}=0.75$; for low positive, $d_{+}=0.25$.
For high negative perturbation we use $f_{-}=0.7,\ d_{-}=0.8$; for low negative, $f_{-}=0.2,\ d_{-}=0.2$.
The canonical setting (v0) fixes $d_{+}=0.75,\ f_{-}=0.7,\ d_{-}=0.8$.
hi-pos/lo-neg is the most challenging regime: the positive undergoes maximal surface disruption while the negative carries only minimal integrity-breaking change, forcing the encoder to rank a heavily permuted but structurally faithful table above a barely corrupted one.
lo-pos/hi-neg is the inverse and the easiest regime.
We select the balanced hi-pos/hi-neg setting with both permutation types as the canonical variation v0, reported in the main paper.

\begin{table}
\centering
\footnotesize
\begin{tabular}{lccccc}
\toprule
& \textbf{\shortstack{hi-pos \\ hi-neg}} & \textbf{\shortstack{lo-pos \\ hi-neg}} & \textbf{\shortstack{hi-pos \\ lo-neg}} & \textbf{\shortstack{BIG \\ (hi-pos/hi-neg)}} & \textbf{\shortstack{SMALL \\ (hi-pos/hi-neg)}} \\
\midrule
both        & v0 & v1 & v2 & v9  & v12 \\
row\_reorder & v3 & v4 & v5 & v10 & v13 \\
col\_reorder & v6 & v7 & v8 & v11 & v14 \\
\bottomrule
\end{tabular}
\caption{Shuffling variation grid: Permutation $\times$ Magnitude (v0--v8) and Permutation $\times$ Size (v9--v14, fixed hi-pos/hi-neg). v0 is the canonical setting.}
\label{tab:shuffling-grid}
\end{table}

Table~\ref{tab:shuffling_magnitude} reports accuracy across magnitude regimes and permutation types. 
Discrimination difficulty increases monotonically with positive perturbation magnitude and decreases with negative perturbation magnitude, because the two axes operate on the same underlying margin. 
Larger positive perturbation drives the positive sample further from the anchor in surface-text space, compressing the gap an encoder must bridge to score it above the negative. 
Conversely, smaller negative perturbation keeps the negative textually close to the anchor, shrinking the margin from the other side. 
Both effects independently tighten the structural signal an encoder must exploit to succeed.

\begin{table}
\centering
\setlength{\tabcolsep}{8.11pt}
\begin{tabular*}{\columnwidth}{lccc|c}
\hline
Approach & \makecell{hi-pos \\ hi-neg} & \makecell{lo-pos \\ hi-neg} & \makecell{hi-pos \\ lo-neg} & Mean \\
\hline
Granite-R2 (md) & 0.1286 & 0.5705 & \underline{0.0091} & 0.2361 \\
GritLM (md) & 0.1802 & \underline{0.8426} & 0.0036 & \underline{0.3422} \\
Hashing & 0.0000 & 0.0000 & 0.0000 & 0.0000 \\
HyTrel & \textbf{0.8747} & \textbf{0.9712} & \textbf{0.8277} & \textbf{0.8912} \\
MiniLM (md) & \underline{0.1965} & 0.6376 & 0.0036 & 0.2792 \\
\hline
\end{tabular*}
\caption{Table Shuffling:  Magnitude grid. Accuracy averaged over all 6 datasets (perturbation=both, default window). Rows = positive-magnitude / negative-magnitude. Best in bold.}
\label{tab:shuffling_magnitude}
\end{table}

\noindent\textbf{Size Ablation.}
The same grid (Table~\ref{tab:shuffling-grid}) crosses permutation type with two extreme table window sizes (BIG and SMALL), fixing hi-pos/hi-neg, where variations v9--v11 use large windows and v12--v14 use small windows.

\begin{table}
\centering
\scriptsize
\begin{tabular*}{\columnwidth}{lcccccc|c}
\hline
Approach & \multicolumn{2}{c}{Both} & \multicolumn{2}{c}{Row reorder} & \multicolumn{2}{c}{Col reorder} & \multicolumn{1}{c}{} \\
 & BIG & SMALL & BIG & SMALL & BIG & SMALL & Mean \\
\hline
Granite-R2 (md) & 0.1856 & 0.2114 & 0.4473 & 0.4507 & 0.4781 & 0.4973 & 0.3784 \\
GritLM (md) & 0.2375 & \underline{0.2619} & \underline{0.5214} & \underline{0.5646} & 0.6040 & 0.5554 & \underline{0.4575} \\
Hashing & 0.0000 & 0.0000 & 0.0000 & 0.0000 & 0.0000 & 0.0000 & 0.0000 \\
HyTrel & \textbf{0.9654} & \textbf{0.9850} & \textbf{0.9995} & \textbf{0.9995} & \textbf{0.9694} & \textbf{0.9824} & \textbf{0.9835} \\
MiniLM (md) & \underline{0.2434} & 0.2345 & 0.3906 & 0.4270 & \underline{0.6821} & \underline{0.6403} & 0.4363 \\
\hline
\end{tabular*}
\caption{Table Shuffling: Perturbation-type breakdown \textbf{and} size ablation (hi-pos/hi-neg). Averaged over fetaqa, tabfact, ottqa, spider-train. CKAN and ECB excluded (no tables pass SMALL window filter).}
\label{tab:shuffling_size}
\end{table}

We test whether structural sensitivity depends on table dimensions by comparing accuracy across BIG and SMALL windows for each permutation type. 
Table~\ref{tab:shuffling_size} shows this robustness holds broadly, with approaches ranking consistent across window sizes. 
One asymmetry stands out: HyTrel's accuracy on column reordering trails row reordering, suggesting that the structure-aware encoder's representations are not completely column-position independent.

\noindent\textbf{Per-dataset Accuracy.}
The v0 setting (hi-pos/hi-neg magnitude, both row+column permutation, default window; Table~\ref{tab:shuffling-grid}) gives one aggregate accuracy per approach, but this could mask dataset-specific failure or success. 
We therefore report the full per-dataset breakdown at v0 in Table~\ref{tab:shuffling_accuracy} to check whether trends hold uniformly or are driven by a subset of corpora. 
HyTrel's advantage holds everywhere, most strikingly on ECB and TabFact where it reaches near-perfect accuracy while every transformer and Hashing score exactly zero.

\begin{table*}[t]
\centering
\setlength{\tabcolsep}{3.84pt}
\begin{tabular*}{\textwidth}{lcccccc|c}
\hline
Approach & ckan\_subset & ecb & fetaqa & ottqa & spider-train & tabfact & Mean \\
\hline
Granite-R2 (csv) & 0.0549 {\scriptsize [0.049, 0.061]} & \underline{0.0000} {\scriptsize [0.000, 0.000]} & 0.2200 {\scriptsize [0.184, 0.256]} & 0.2285 {\scriptsize [0.192, 0.267]} & 0.1344 {\scriptsize [0.104, 0.165]} & 0.1342 {\scriptsize [0.118, 0.151]} & 0.1287 \\
Granite-R2 (md) & 0.0585 {\scriptsize [0.052, 0.065]} & \underline{0.0000} {\scriptsize [0.000, 0.000]} & 0.2340 {\scriptsize [0.196, 0.272]} & 0.1984 {\scriptsize [0.164, 0.234]} & 0.1222 {\scriptsize [0.094, 0.153]} & 0.1584 {\scriptsize [0.141, 0.176]} & 0.1286 \\
GritLM (csv) & 0.1719 {\scriptsize [0.161, 0.182]} & \underline{0.0000} {\scriptsize [0.000, 0.000]} & 0.3160 {\scriptsize [0.276, 0.356]} & 0.2545 {\scriptsize [0.216, 0.293]} & 0.1141 {\scriptsize [0.088, 0.143]} & 0.1348 {\scriptsize [0.118, 0.151]} & 0.1652 \\
GritLM (md) & 0.1585 {\scriptsize [0.149, 0.169]} & \underline{0.0000} {\scriptsize [0.000, 0.000]} & \underline{0.3680} {\scriptsize [0.326, 0.412]} & 0.2766 {\scriptsize [0.238, 0.317]} & 0.1181 {\scriptsize [0.090, 0.147]} & 0.1602 {\scriptsize [0.143, 0.178]} & 0.1802 \\
Hashing & 0.0000 {\scriptsize [0.000, 0.000]} & \underline{0.0000} {\scriptsize [0.000, 0.000]} & 0.0000 {\scriptsize [0.000, 0.000]} & 0.0000 {\scriptsize [0.000, 0.000]} & 0.0000 {\scriptsize [0.000, 0.000]} & 0.0000 {\scriptsize [0.000, 0.000]} & 0.0000 \\
HyTrel & \textbf{0.3824} {\scriptsize [0.369, 0.396]} & \textbf{1.0000} {\scriptsize [1.000, 1.000]} & \textbf{0.8740} {\scriptsize [0.844, 0.902]} & \textbf{0.9940} {\scriptsize [0.986, 1.000]} & \textbf{0.9980} {\scriptsize [0.994, 1.000]} & \textbf{1.0000} {\scriptsize [1.000, 1.000]} & \textbf{0.8747} \\
MiniLM (csv) & 0.1862 {\scriptsize [0.176, 0.197]} & \underline{0.0000} {\scriptsize [0.000, 0.000]} & 0.2380 {\scriptsize [0.202, 0.276]} & 0.2104 {\scriptsize [0.174, 0.248]} & 0.1446 {\scriptsize [0.114, 0.177]} & 0.1507 {\scriptsize [0.134, 0.167]} & 0.1550 \\
MiniLM (md) & \underline{0.2160} {\scriptsize [0.205, 0.227]} & \underline{0.0000} {\scriptsize [0.000, 0.000]} & 0.2820 {\scriptsize [0.242, 0.322]} & \underline{0.3086} {\scriptsize [0.269, 0.349]} & \underline{0.1629} {\scriptsize [0.132, 0.196]} & \underline{0.2092} {\scriptsize [0.190, 0.229]} & \underline{0.1965} \\
\hline
\end{tabular*}
\caption{Table Shuffling: Accuracy per dataset (v0: hi-pos/hi-neg, both row+col perturbation, default window). Best per column in bold, second-best underlined. Bracketed values show bootstrapped 95\% CIs.}
\label{tab:shuffling_accuracy}
\end{table*}

\begin{figure*}
\centering
\includegraphics[width=0.7\textwidth]{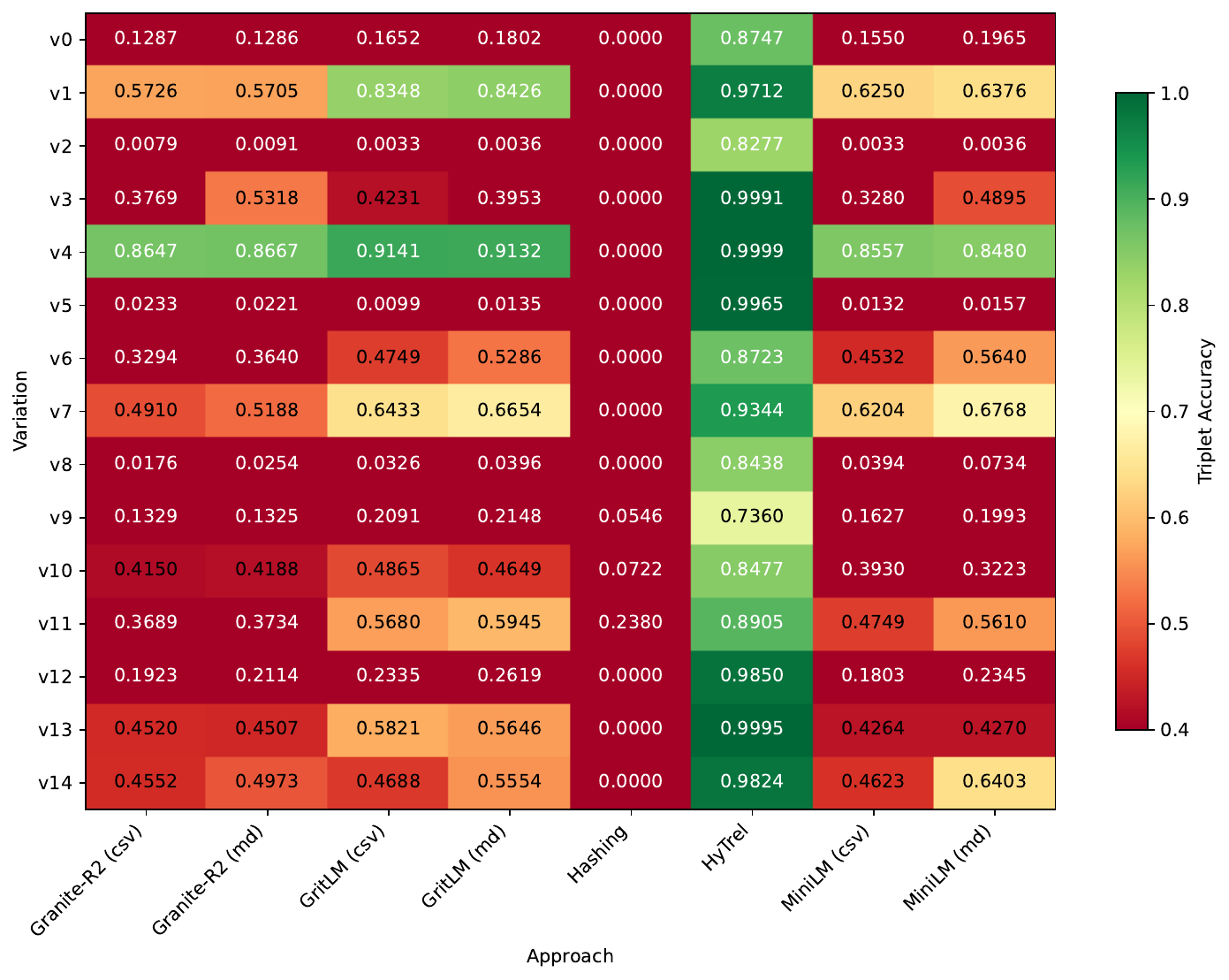}
\caption{Table Shuffling: Accuracy heatmap across all variations v0--v14 (defined in Table \ref{tab:shuffling-grid}) and approaches, averaged over datasets. The structure-aware encoder maintains uniformly high accuracy; transformer columns show variation-dependent structural sensitivity.}
\label{fig:variation-heatmap}
\end{figure*}

\noindent\textbf{Full Grid Overview.}
Having confirmed per-dataset consistency at v0, we check whether these trends extend across all 15 variations rather than just the canonical setting; the results are visualized in Figure~\ref{fig:variation-heatmap}.
HyTrel~\cite{chen2023hytrel} maintains uniformly high accuracy throughout the grid, while transformer columns exhibit variation-dependent sensitivity with performance dropping in higher perturbation regimes, and the hashing approach predictably scores zero as the extreme failure mode of this task.

\begin{figure}
\centering
\includegraphics[width=\columnwidth]{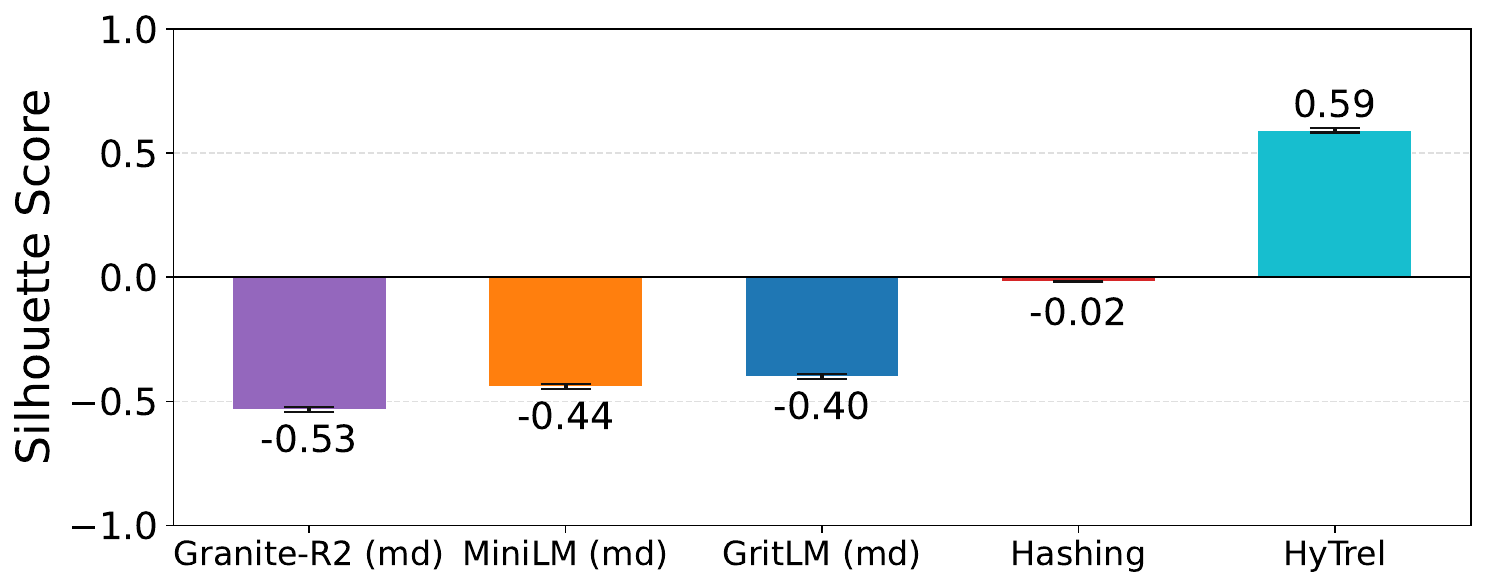}
\caption{Table Shuffling: Silhouette score per approach, averaged across all datasets at the v0 variation (pos\_type=both, hi-pos/hi-neg, markdown serialization). Error bars show bootstrapped 95\% CIs, pooled across datasets. Hashing's near-zero score reflects its permutation invariance: the embedding cannot distinguish the anchor, positive, and negative as different vectors, confirming the evaluation protocol does not reward bag-of-words representations.}
\label{fig:silhouette-bars}
\end{figure}

\noindent\textbf{Silhouette Score.}
Accuracy alone doesn't reveal how confidently an approach separates positives from negatives, so we compute the silhouette score (Eq. 1) per approach, averaged across datasets at v0. 
Figure~\ref{fig:silhouette-bars} shows Hashing sits at exactly zero, reflecting its permutation invariance rather than weak discrimination — it cannot represent anchor, positive, and negative as distinct vectors at all. 
The transformers score negative, indicating systematic inversion rather than merely weak signal, while HyTrel is the only approach with a clearly positive margin.

\begin{figure}
\centering
\includegraphics[width=0.8\columnwidth]{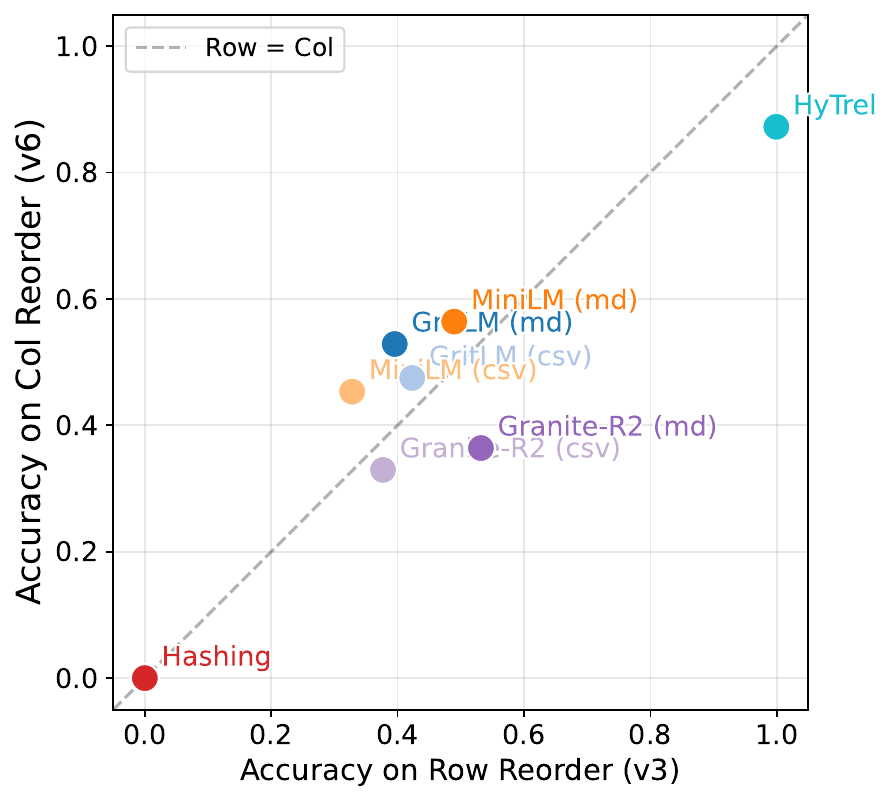}
\caption{Table Shuffling: Row vs.\ column perturbation accuracy (hi-pos/hi-neg, default table window, averaged over datasets). Points above the diagonal are column-sensitive; points below are row-sensitive. HyTrel sits near the top-right corner, indicating slightly weaker robustness to column than row reordering. Transformer points spread off-diagonal, revealing encoder-specific asymmetries.}
\label{fig:row-col-scatter}
\end{figure}

\noindent\textbf{Row vs. Column Sensitivity.}
Aggregate accuracy conflates row and column reordering, which could affect encoders differently. 
Figure~\ref{fig:row-col-scatter} compares per-approach accuracy on row-only (v3) vs. column-only (v6) variations: HyTrel sits near the top-right corner but slightly below the diagonal, indicating mildly weaker robustness to column reordering, while transformers scatter off-diagonal, each showing its own directional asymmetry.

\subsection{Table Type Detection}
\label{sec:appendix-ttd}

\noindent\textbf{Per-Classifier Results.}
Section~\ref{sec:ttd-results} reports macro-F1 aggregated across classifiers, which could mask whether rankings depend on the probe used. Table~\ref{tab:ttd_classifier} breaks down accuracy and macro-F1 by classifier. The dense transformer approaches (GritLM, Granite-R2, MiniLM) rank consistently relative to one another across XGBoost, MLP, and KNeighbors, but Hashing shows a pronounced classifier dependence: it leads under XGBoost yet trails all transformer encoders under MLP and KNeighbors, consistent with its bag-of-words representation being well-suited to tree-based decision boundaries but less effective for neural or distance-based classifiers. HyTrel ranks last across all three probes.

\begin{table}[H]
\centering
\scriptsize
\setlength{\tabcolsep}{3.7pt}
\begin{tabular*}{\columnwidth}{lcccccc}
\hline
Approach & \multicolumn{1}{c}{XGBoost} & \multicolumn{1}{c}{XGBoost} & \multicolumn{1}{c}{MLP} & \multicolumn{1}{c}{MLP} & \multicolumn{1}{c}{KNeighbors} & \multicolumn{1}{c}{KNeighbors} \\
 & accuracy & macro-F1 & accuracy & macro-F1 & accuracy & macro-F1 \\
\hline
Granite-R2 (csv) & 0.8980 & 0.8959 & 0.9090 & 0.9078 & 0.8560 & 0.8507 \\
Granite-R2 (md) & \underline{0.9110} & \underline{0.9103} & 0.9240 & 0.9233 & 0.8600 & 0.8549 \\
GritLM (csv) & 0.9090 & 0.9083 & \underline{0.9330} & \underline{0.9324} & \underline{0.8690} & \underline{0.8627} \\
GritLM (md) & 0.9060 & 0.9059 & \textbf{0.9380} & \textbf{0.9375} & \textbf{0.8820} & \textbf{0.8758} \\
Hashing & \textbf{0.9358} & \textbf{0.9252} & 0.8684 & 0.8503 & 0.8376 & 0.8017 \\
HyTrel & 0.6490 & 0.6439 & 0.6610 & 0.6557 & 0.5630 & 0.5545 \\
MiniLM (csv) & 0.7910 & 0.7888 & 0.8100 & 0.8070 & 0.7680 & 0.7519 \\
MiniLM (md) & 0.7850 & 0.7818 & 0.8180 & 0.8154 & 0.7600 & 0.7443 \\
\hline
\end{tabular*}
\caption{Table Type Detection: Accuracy and macro-F1 per classifier on WDC Schema.org (header-stripped, frozen embeddings). Best per column in bold, second-best underlined.}
\label{tab:ttd_classifier}
\end{table}

\subsection{Cross-Task Results}
\label{sec:cross-task-results}

\begin{figure}
\centering
\includegraphics[width=\columnwidth]{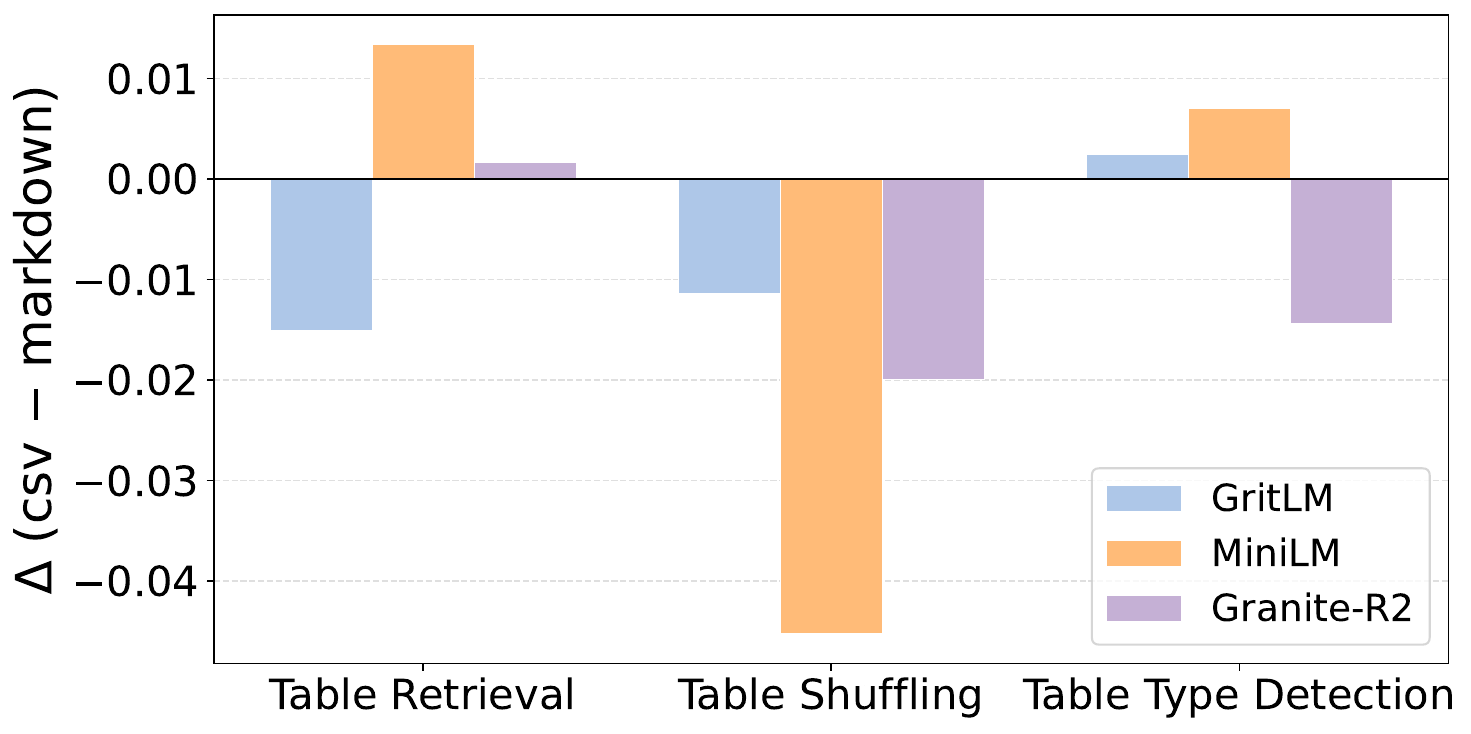}
\caption{Serialization deltas ($\Delta = \text{CSV} - \text{markdown}$; positive means CSV scores higher) across tasks for transformer approaches. Sign and magnitude vary across the approach $\times$ task grid.}
\label{fig:serialization-deltas}
\end{figure}

\noindent\textbf{Serialization Effects Across Tasks.}
Section~\ref{sec:retrieval-results} analyzed serialization effects for table retrieval; here we extend the comparison across all three tasks.
Figure~\ref{fig:serialization-deltas} confirms that the sign and magnitude of the delta between CSV and markdown differ across the \mbox{approach $\times$ task} grid, with no format uniformly preferable, though shuffling shows the largest swings, especially for MiniLM~\cite{wang2020minilm}.

\noindent\textbf{Overall Ranking.}
Section~\ref{sec:cross-task-overview} argues no single approach dominates all axes; Table~\ref{tab:overall_ranking} makes this concrete by ranking each approach per task and averaging. 
GritLM has the best average rank overall, but no approach ranks first on more than one task, reinforcing that per-task and aggregate rankings disagree.

\begin{table}
\centering
\setlength{\tabcolsep}{3pt}
\begin{tabular*}{\columnwidth}{lcccc}
\hline
Approach & \makecell{Table \\ Retrieval \\ (MRR@10)} & \makecell{Table \\ Shuffling \\ (Accuracy)} & \makecell{Table Type \\ Detection \\ (XGB F1)} & Overall \\
\hline
GritLM (md) & \textbf{1.00} & 3.00 & 3.00 & \textbf{2.33} \\
Granite-R2 (md) & \underline{2.00} & 4.00 & \underline{2.00} & \underline{2.67} \\
MiniLM (md) & 3.00 & \underline{2.00} & 4.00 & 3.00 \\
Hashing & 4.00 & 5.00 & \textbf{1.00} & 3.33 \\
HyTrel & 5.00 & \textbf{1.00} & 5.00 & 3.67 \\
\hline
\end{tabular*}
\caption{Per-task and overall rank of each approach (1=best, 5=worst). No approach ranks first on more than one task, underscoring that no single embedding has all properties.}
\vspace{-1em}
\label{tab:overall_ranking}
\end{table}

\begin{figure}
\centering
\includegraphics[width=0.8\columnwidth]{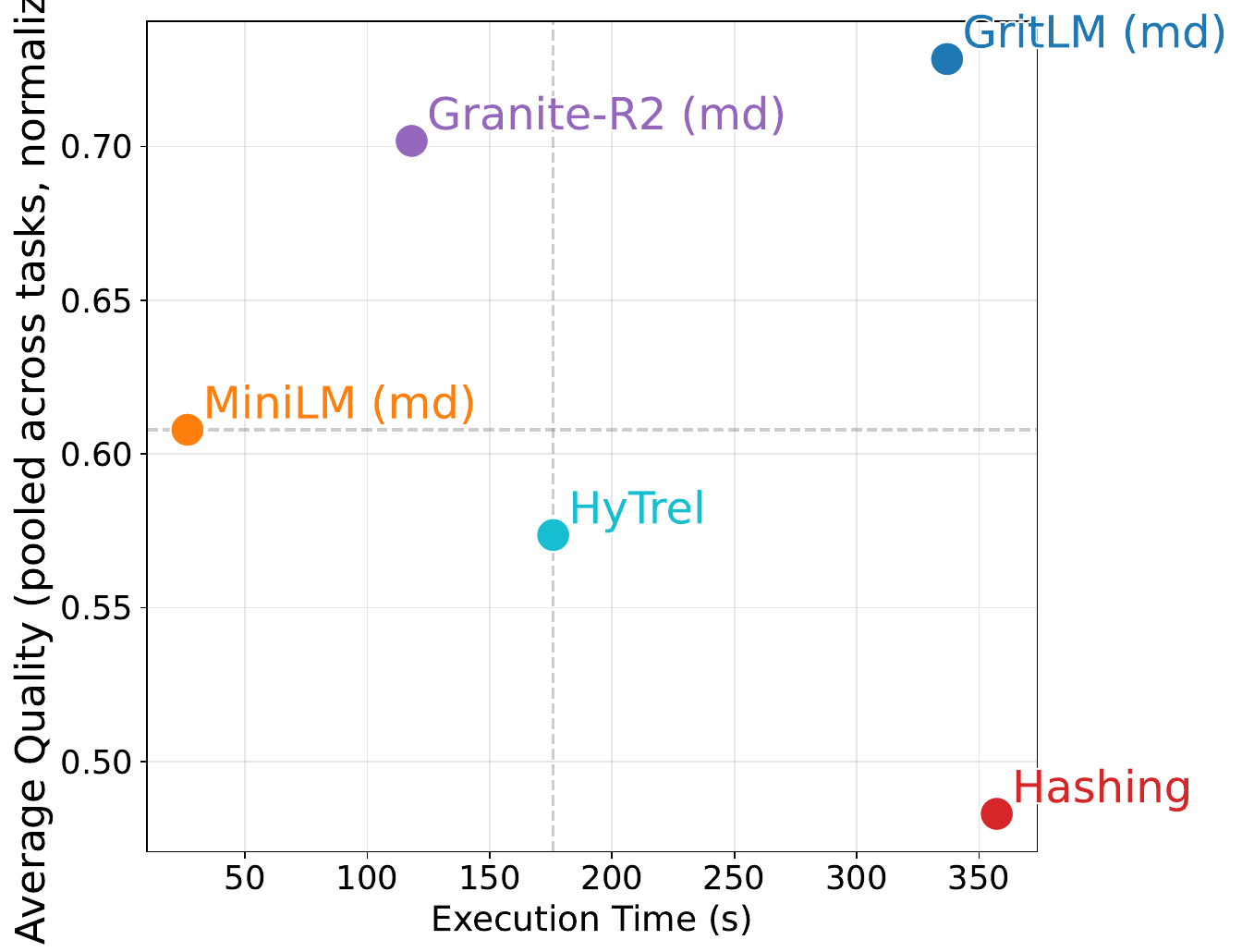}
\caption{Cost--quality trade-off: Quality is the mean of each approach's three per-task scores, cost is the total embedding time in seconds.}
\label{fig:cost-quality}
\end{figure}

\noindent\textbf{Cost-Quality Trade-off.}
Beyond raw quality, practical model selection depends on embedding cost. 
Figure~\ref{fig:cost-quality} relates pooled quality to embedding time: cost and quality are not consistently related.
GritLM's higher cost does pay off in higher quality, but Hashing incurs high cost for the lowest quality (due to its 32,768-dimensional sparse vectors), and Granite-R2 offers the best cost--quality trade-off in the benchmark, indicating quality is driven more by architecture and training than by compute spent at inference.

\end{document}